%% file: master-sigconf.tex
\documentclass[sigconf]{acmart}
\usepackage{booktabs} % For formal tables
\usepackage{listings}
\usepackage{subcaption}
\usepackage{multirow}
\usepackage{balance}

\setcopyright{acmcopyright}
\acmConference[CASCON17]{27th Annual International Conference on Computer Science and Software Engineering}
{November 2017}{Toronto, Ontario, Canada} 
\acmYear{2017}
\copyrightyear{2017}
\acmPrice{15.00}

\begin{document}
\title{Elascale: Autoscaling and Monitoring as a Service}
% \titlenote{Produces the permission block, and
% \subtitle{Extended Abstract}
% \subtitlenote{The full version of the author's guide is available as
%   \texttt{acmart.pdf} document}

\author{Hamzeh Khazaei, Rajsimman Ravichandiran, Byungchul Park, Hadi Bannazadeh, Ali Tizghadam and 
Alberto Leon-Garcia}
% \authornote{Dr.~Trovato insisted his name be first.}
\affiliation{%
  \institution{Department of Electrical and Computer Engineering, University of Toronto}
%   \streetaddress{University of Toronto}
  \city{Toronto} 
  \state{ON, Canada} 
%    \postcode{43017-6221}
}
\email{{firstname.lastname}@utoronto.ca}

% \authornote{The secretary disavows any knowledge of this author's actions.}
% \affiliation{%

% \authornote{This author is the
%   one who did all the really hard work.}
% \affiliation{%

% \author{Lawrence P. Leipuner}
% \affiliation{
%   \institution{Brookhaven Laboratories}
%   \streetaddress{P.O. Box 5000}}
% \email{lleipuner@researchlabs.org}

% \author{Sean Fogarty}
% \affiliation{%
%   \institution{NASA Ames Research Center}
%   \city{Moffett Field}
%   \state{California} 
%   \postcode{94035}}
% \email{fogartys@amesres.org}

% \affiliation{%


% \author{Julius P.~Kumquat}

\renewcommand{\shortauthors}{H. Khazaei et al.}

\begin{abstract}
Auto-scalability has become an evident feature for cloud software systems including but not limited to big data and IoT 
applications. Cloud application providers now are in full control over their applications' microservices and macroservices; 
virtual machines and containers can be provisioned or deprovisioned on demand at runtime. Elascale strives to adjust both micro/macro 
resources with respect to workload and changes in the internal state of the whole application stack. Elascale leverages 
Elasticsearch stack for collection, analysis and storage of performance metrics. Elascale then uses its default scaling engine 
to elastically adapt the managed application. Extendibility is guaranteed through provider, schema, plug-in and policy 
elements in the Elascale by which flexible scalability algorithms, including both reactive and proactive techniques, can 
be designed and implemented for various technologies, infrastructures and software stacks. In this paper, we present the 
architecture and initial implementation of Elascale; an instance will be leveraged to add auto-scalability to a generic
IoT application. Due to zero dependency to the target software system, Elascale can be leveraged to provide 
auto-scalability and monitoring as-a-service for any type of cloud software system.
\end{abstract}

\begin{CCSXML}
<ccs2012>
<concept>
<concept_id>10011007.10010940.10010971.10011120.10003100</concept_id>
<concept_desc>Software and its engineering~Cloud computing</concept_desc>
<concept_significance>500</concept_significance>
</concept>
<concept>
<concept_id>10011007.10010940.10011003.10011002</concept_id>
<concept_desc>Software and its engineering~Software performance</concept_desc>
<concept_significance>500</concept_significance>
</concept>
</ccs2012>
\end{CCSXML}

\ccsdesc[500]{Software and its engineering~Cloud computing}
\ccsdesc[500]{Software and its engineering~Software performance}


% We no longer use \terms command

\keywords{Auto-scalability, monitoring, microservices, macroservices, containers,
docker, elasticsearch, scalability as a service, cloud application.}

%% Used in some conference proceedings e.g. sigplan and sigchi
%   \caption{This is a teaser}

\maketitle

\input{body-conf.tex}

\bibliographystyle{ACM-Reference-Format}
\bibliography{sigproc} 

\end{document}

%% file: body-conf.tex
\section{Introduction} \label{intro}
Cloud elasticity and software defined paradigm are the enabling building blocks for 
\emph{auto-scalability} which is one of the required and vital features for nowadays cloud software systems 
including infrastructures, platforms and applications. Due to dynamic nature of cloud environments,
workloads, and internal states, cloud software systems needs to constantly adapt themselves to the new 
conditions to maintaining their service level agreements (SLA) while utilizing their resources efficiently. 

\begin{figure}[!t]
\includegraphics[width=0.99\columnwidth]{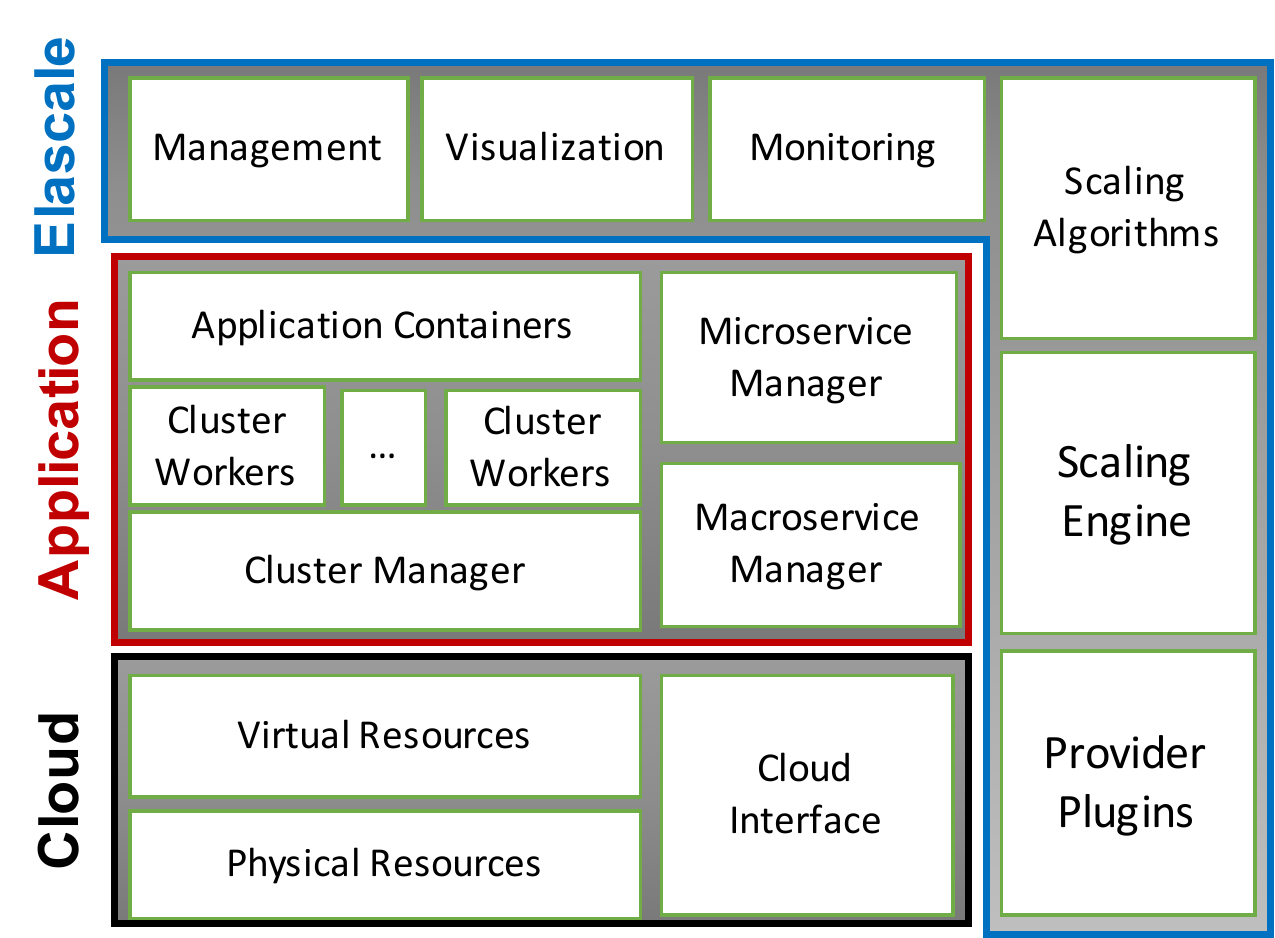}
\caption{Detailed View of Elascale.}
\label{fig:elascale-2}
\end{figure}

Recently, a pattern has been adopted by many software-as-a-service providers in which both virtual machines 
(VMs) and containers
are leveraged to offer their solutions as microservices. This way, strong isolation, preferred for higher order 
of security, inherited from hypervisor-based virtualization will be coupled with flexibility and portability of 
containers to offer reliable and easy-to-manage services. Scaling algorithms and solutions in such scenarios are 
required to provide scalability for both VMs and containers. 

So far, we have seen customized and vendor-specific auto-scalability features by cloud service providers 
such as IBM~\cite{bluemix-auto}, Amazon~\cite{amazon-auto} and others. General purpose solutions in the 
open source community, as we survey them in Section~\ref{rw}, often deal with either microservices 
(i.e., containers) or macroservices (i.e., VMs) rather than the whole stack of the application. Such shortcomings 
motivated us to design and implement a general-purpose, cloud and application agnostic monitoring and 
auto-scaling solution for nowadays cloud software systems, which embodies both micro and macro services.

The name \emph{Elascale} is inspired from Elasticsearch~\cite{elasticsearch} as Elascale leverages ELKB  
family including \underline{E}lasticsearch, \underline{L}ogstash, \underline{K}ibana and 
\underline{B}eats~\cite{beats} for the 
monitoring, data processing/storage and visualization purposes. Also, Elascale leverages a core scalability engine
that implements the monitor-analyze-plan-execute-knowledge, i.e., MAPE-K loop, introduced by
IBM~\cite{IBM:2005:ArchitecturalBlueprint}. Elascale automatically instrument the cloud application, for collecting
performance metrics at all layers, and then applies a default threshold-based, reactive scalability algorithm for 
all application's micro and macro services. It discovers all services automatically, and make the user able to set 
values for 
scaling algorithm, e.g., thresholds, scaling steps, cooling time and the like. Extendability, including supporting
various applications, cloud service providers, scaling algorithms, visualization and etc, has been the primary concerns
in designing and implementing Elascale solution. Figure~\ref{fig:elascale-2} shows the high level architecture of 
the Elascale.

This paper is organized as follows; in Section~\ref{bw} present the background concepts and technologies for our
work in this paper. In Section ~\ref{elascale} we present Elascale in details. Section~\ref{ee} presents the 
experimental evaluation and Section ~\ref{rw} surveys related work and projects. In the end, Section~\ref{con} 
concludes the paper and highlights the immediate future work for Elascale.

%==========================================================
\section{Background} \label{bw}
In this section we briefly describe the main related concepts which helps us to present Elascale effectively.

\subsection{Microservices}
Microservices are becoming the main trend in cloud-based application development~\cite{turnbull2014docker}. 
The traditional monolithic application
is decomposed into small pieces that provide a single service: the full capabilities of the application emerge from the 
interaction of these small pieces. Microservices are independent from each other and organized around capabilities, e.g. 
user interface, front-end, etc. Their decoupling allows developers to use the best technology for their implementation
according to the task they have to accomplish: the application becomes polyglot, involving different programming 
languages and technologies. An application composed of microservices is inherently distributed,
being divided into hundreds of different microservices, deployed in a large network infrastructure, that communicate
both in a synchronous or asynchronous way, using REST or a message-based system respectively. Through the
use of microservices, the application can scale efficiently as it is possible to scale only the microservices that 
are under heavy load, not the entire application. The microservices architecture embodies the principles of the 
DevOps (a clipped compound of ``development'' and ``operations'') movement, promoting the automation of deployment 
and testing and reducing the burden on management and operations~\cite{florio2016gru}. 

\subsection{Docker}
Microservices usually run into containers like Docker~\cite{turnbull2014docker, bernstein2014containers}. 
Docker containers can run an application as
an isolated process on a host machine, including only the application and all its dependencies (the kernel of the Operating
System is shared among other containers) and providing to it only the resources it requires. Docker containers are 
different from a fully virtualized system like a virtual machine: a virtual machine contains a full OS that runs 
in isolation on physical
resources that are virtualized by an hypervisor on the basis of the ones available in the host machine; a Docker 
container uses the resources available in the host (both a physical or a virtualized one) that are assigned to it 
by the Docker Engine.
The consequence is that Docker containers can share physical resources and are lightweight: it is possible to run multiple
containers on the same machine starting them in seconds. Docker allows developers to implement their application and
their services using the technology or language that is most suitable to them. Services deployed in a Docker container can
be scaled or replaced just starting or stopping the container running that specific service. Moreover Docker containers can
be deployed in very different settings, from servers in a cloud computing infrastructure to ARM-based IoT devices.

\subsubsection{Docker Swarm}
Distributed applications need distributed system and compute resources on it. Docker Swarm is a clustering and scheduling
tool, which offers functionalities to turn a group of Docker Systems (Nodes) into a Virtual Docker System. It builds
a cooperative group of systems that can provide redundancy if one or more nodes fail. Swarm provides workload balancing
for containers~\cite{swarm}. It assigns containers to underlying nodes and optimizes resources by automatically 
scheduling container workloads to run on the most appropriate host with adequate resources while maintaining necessary 
performance levels ~\cite{naik2016building}. An IT administrator or developer controls Swarm using a swarm manager, 
which organizes and schedules containers. Kubernetes~\cite{kubernetes} is an alternative for Docker Swarm that is introduced 
by Google. 

\subsubsection{Docker Machine}
Docker Machine is a tool that makes it easy to provision and manage multiple Docker hosts remotely 
from a personal computer. Such servers are commonly referred to as Dockerized hosts that can be used 
to run Docker containers. Docker Machine supports various
backend cloud service providers such as Amazon Web Services, Microsoft Azure, Digital Ocean, Google Compute
Engine, Exoscale, Generic, OpenStack, Rackspace, IBM Softlayer and VMware vCloud Air which makes it ideal to
manage the macroservices centrally regardless of their location and vendor~\cite{docker-machine}. Docker 
Machine in combination with Docker Swarm can be used to build a virtual system of systems in multiple
clouds~\cite{naik2016building}.

\subsection{ELK and Beats}
Beats are lightweight agents that can send performance metrics data from containers, VMs or large number 
of software products to Logstash or Elasticsearch~\cite{beats}. Currently, there exist more than 70 different 
Beats, developed by Elastic and the community in the Beats family, including Dockbeat and Metricbeat that have
been used in the initial version of Elascale. Logstash is a dynamic 
data collection pipeline with an extensible plugin ecosystem and strong Elasticsearch synergy. 
Elasticsearch is a distributed, JSON-based search and analytics engine designed for horizontal scalability, 
maximum reliability, and easy management. Kibana gives shape to the data and is the extensible user 
interface for configuring and managing all aspects of the Elastic Stack~\cite{elasticsearch}. 

%==========================================================
\section{Elascale} \label{elascale}

\begin{figure}
\centering
\begin{subfigure}{0.5\columnwidth}
  \centering
  \includegraphics[width=0.9\columnwidth]{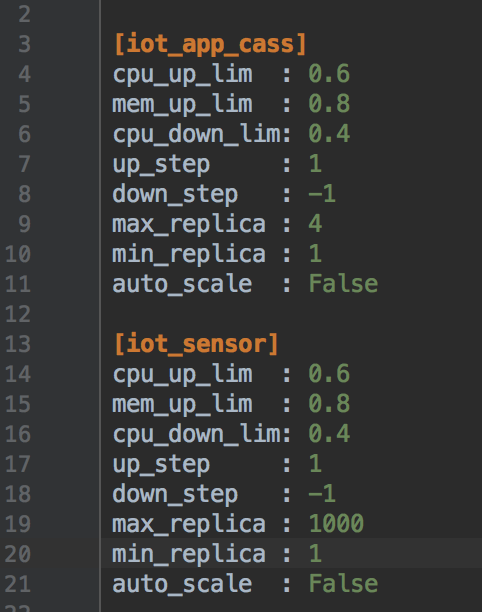}
  \caption{microservice.ini}
  \label{fig:micro-ini}
\end{subfigure}%
\begin{subfigure}{0.5\columnwidth}
  \centering
  \includegraphics[width=0.96\columnwidth]{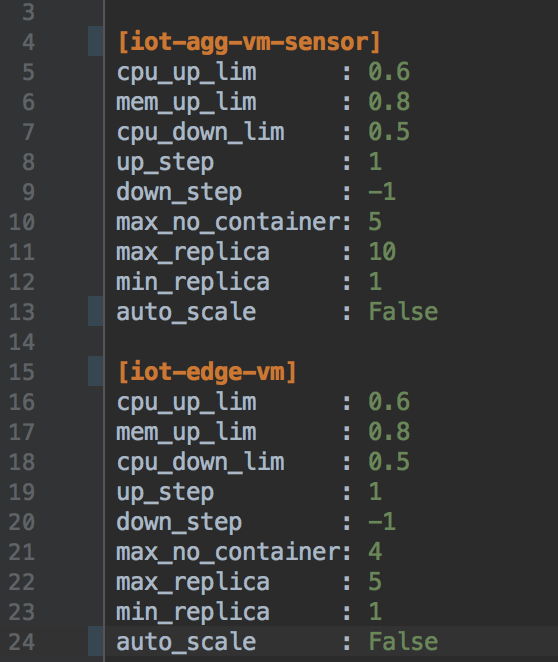}
  \caption{macroservice.ini}
  \label{fig:macro-ini}
\end{subfigure}
\caption{Excerpts of the configuration files for microservices and macroservices created by Elascale.}
\label{fig:ini}
\end{figure}

We now present our approach, \emph{Elascale\footnote{\url{https://gitlab.com/hamzeh.khazaei/Elascale}}}. 
The two important components of Elascale are: (i) an ELKB based
monitoring stack, and (ii) an auto-scaling engine. Together, the two 
component provides auto-scalability for any cloud application that has been deployed through microservices.
Currently, Elascale purely supports Docker technology family and the plan is to support non-docker solution
in next versions, Kubernetes for clustering in particular. Therefore, Elascale V0.9 (current version) assumes 
and leverages the followings:
\begin{itemize}
\item Docker as the container engine
\item Application has been deployed through `service' or `stack' commands
\item Docker Swarm for clustering
\item Docker Machine to manage the backend cloud(s)
\end{itemize}
Elascale itself is deployed as a microservice on Swarm Master node or another machine that has
access to the Docker engine on the Swarm Master node. The following command deploys Elascale as a microservice
on Swarm Master node which is set to scale both microservices and macroservices for the current application:
\begin{lstlisting} [label={lst:elascale},caption={A sample deployment of Elascale}]
$ docker service create --name elascale --env MICRO=TRUE
   --env MACRO=TRUE --env CLUSTER=SWARM  
   --env DOCKER_MACHINE=TRUE
   --constraint 'node.labels.role==manager'    
   henaras/elascale:0.9
\end{lstlisting}
The above command triggers scalability for both microservices and macroservices, could be false for each, and assumes
Swarm as the cluster management system. Also it assumes that Docker Machine has already been installed and configured. 
Now we dive down under the hood to see what will happen after issuing above command:
\begin{enumerate}

\item First, it will create a virtual machine and then deploy Elasticsearch, Logstash and Kibana as ``\emph{replicated}'' 
docker services. 

\item Elascale deploys Metricbeat and Dockbeat on all VMs using a ``\emph{global}'' docker service; this way any newly 
added node to the Swarm cluster will get these two beats automatically. 

\item Elascale then discovers all microservices and create ``\emph{microservice.ini}'' that contains the list of 
application microservices along with default parameters to be used by the Elascale's scaling engine.
Figure~\ref{fig:micro-ini} shows an excerpt of this file for our sample IoT application. As can be seen, 
auto scalability is disabled by default
for services; at this time a web user interface will be generated out of this file for the user to be customized.
The user, i.e., the application owner can change parameters to suit their needs.

\item With the same fashion, Elascale discovers macroservices, i.e., virtual machines, and create
``\emph{macroservice.ini}'' file to be customized by the user in the Elascale web UI, like microservices.
Figure~\ref{fig:macro-ini} shows an excerpt of this file for our sample IoT application.

\item After customizing the configuration files by the application owner, the Elascale auto-scaling engine starts 
monitoring the application's microservices and macroservices based on the active scaling algorithm.
\end{enumerate}

\begin{figure}[!t]
\includegraphics[width=0.95\columnwidth]{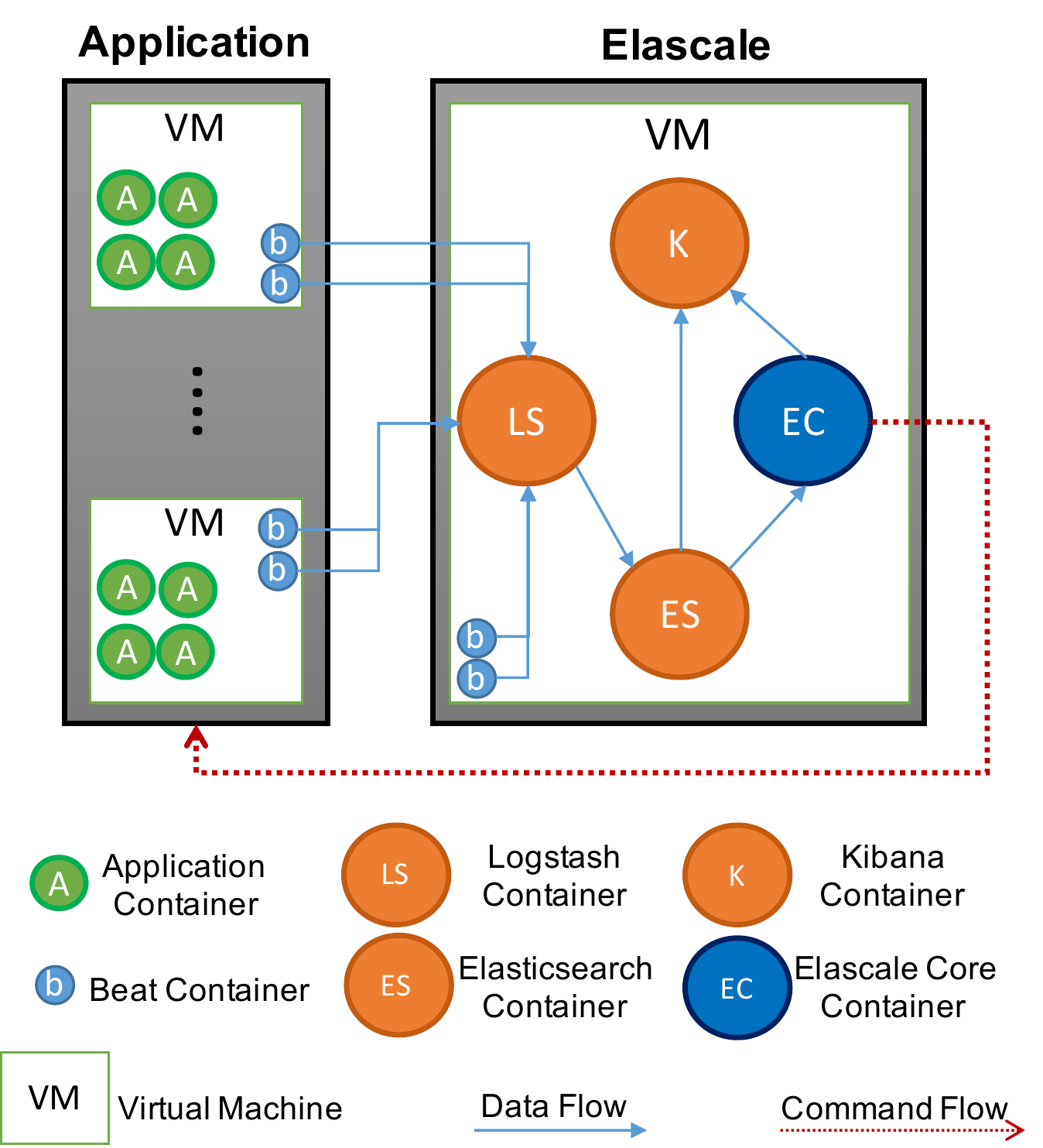}
\caption{Implementation View of Elascale.}
\label{fig:elascale-3}
\end{figure}

The default scaling algorithm in Elascale uses different criteria to scale out/in services. We adopt function
$f$ as the generic formula to scale the micro and macro services.
\begin{multline}
f = \alpha \cdot cpu_{util} + \beta \cdot mem_{util} + \\
\gamma \cdot net_{util} + \lambda \cdot \frac{rep_{fac}^{t}}{rep_{fac}^{c}}
\end{multline}

\noindent in which $\alpha + \beta + \gamma + \lambda = 1$.
In addition to cpu, memory, and network utilization (percentages \%), we incorporate replication 
factor (i.e., $rep_{fac}$) for services which is the dependency factor among services. 
For example, defining the target replication factor (i.e., $rep_{fac}^{t}$) 
of service `$X$' to service `$Y$' as 50\% means that for each instance of service $X$, i.e., for each container, 
there should be at least 2 containers of $Y$ up and running. The main idea of using the ratio of target replication factor
to current replication factor (i.e., $rep_{fac}^{c}$) is to maintain QoS in times of nodes' failure 
or network partitioning. In other words, if due to an internal failure some of services become 
down or unreachable, Elascale redeploys missing services autonomically regardless of  
resource utilizations. The weights of parameters (i.e., $\alpha, \beta, \gamma, \lambda$) can be
tuned based on a bottleneck analysis that is performed during application test. 
By assigning more weight to $\lambda$, we can guarantee a higher level of reliability for our 
application. The details of the default algorithm has been elaborated in~\cite{khazaei-ficloud-2017}. 
Figure~\ref{fig:elascale-3} shows the data flow and command flow between target application and Elascale.

%==========================================================
\section{Experimental Evaluation} \label{ee}
In this section, we elaborate on our experimental setup as well as results. 
In the experiment, we use the latest stable versions of all software components, as of June
2017, in Elascale and the sample IoT application.

%----------------------------------------------------------------------------------------
\subsection{Experimental Setup} \label{es}
We deployed a sample IoT application using SAVI-IoT~\cite{khazaei-ficloud-2017} platform on SAVI 
Cloud~\cite{savi}. Our application leverages the SAVI Core-Cloud at the University of Toronto and 
one of the SAVI edges located at the University of Victoria. Using SAVI-IoT platform, we create 
required VMs on the Core and the Edge cloud; this process includes provisioning of VMs and installing 
Docker packages (i.e., Docker engine and Swarm). Then, a Swarm cluster is created out of provisioned VMs; the
Swarm master will be located at the Core-Cloud. All VMs are labeled and tagged with their roles 
and locations. Next, the microservices will be deployed on top of macroservices at all layers; 
related microservices will be linked to constitute the application logic by the SAVI-IoT platform. 

\begin{figure}[!t]
\includegraphics[width=0.95\columnwidth]{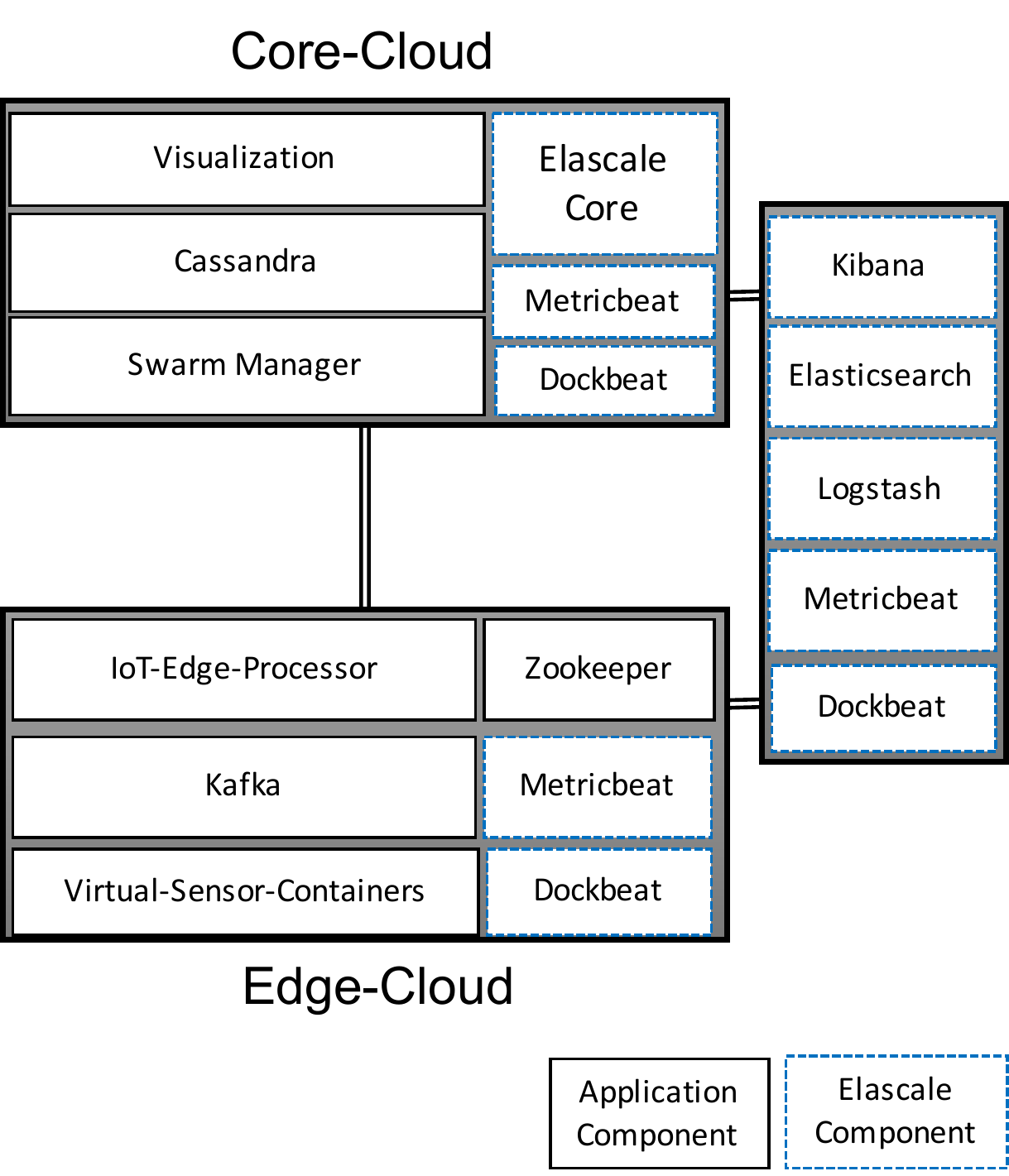}
\caption{Orchestration of Sample IoT application and Elascale components.}
\label{fig:iot-exp}
\end{figure}

In the sample IoT application, as sensors, we deploy containerized virtual sensors that 
collect performance metrics including, CPU utilization, network load and memory consumption of 
themselves. We refer to these containers as ``\emph{virtual-sensor-container}''. In other 
words, each virtual-sensor-container embodies three probe sensors that report resource utilizations
every 15 seconds. Every virtual-sensor-container will be attached to a
co-located aggregator, i.e., a microservice that is running Kafka (kafka.apache.org), automatically. 
The Kafka service 
on that aggregator takes the
responsibility to forward the aggregated data from virtual sensors to the upper service. Here we
set the Kafka service to aggregate sensor data for every 60 seconds and then send them up to a service
named IoT-Edge-Processor. This streaming service at the Edge-Cloud aggregates all the received 
data streams from it's aggregators and ingest them into the Cassandra datastore (cassandra.apache.org) 
located at 
the Core-Cloud. Figure~\ref{fig:iot-exp} shows the experimental setup. 

We examine a normal shape workload to evaluate both out-scaling and in-scaling of the 
application. A client program requests for new sensors according to a Poisson process.
Each request is translated to a virtual-sensor-container which embodies 3 virtual sensors
for measuring cpu, memory and network load. The new virtual-sensor-container will be
attached to an aggregator automatically. By adding more and more virtual sensors, the
resource utilization at aggregators gets increased. If it reaches the upper threshold, 
here set to 70\%, according to function $f_{agg}$, Elascale scales out the aggregator 
microservice.
\begin{multline} \label{eq:agg}
f_{agg} = 0.5 \; cpu_{util} + 0.1 \; mem_{util} + \\
0.1 \; net_{util} + 0.3 \; \frac{rep_{fac}^{t}}{rep_{fac}^{c}}
\end{multline}

\noindent Eq.~\ref{eq:agg} reveals that the aggregator service (i.e., Kafka) is cpu 
intensive as the weight of $\alpha$ is equal to all others combined. Here we set the replication factor
of Kafka service relative to virtual-sensor-containers. We use $f_{edge}$
and $f_{core}$ functions for the Edge and Core services.
\begin{multline} \label{eq:edge}
f_{edge} = 0.2 \; cpu_{util} + 0.5 \; mem_{util} + \\
0.1 \; net_{util} + 0.2 \; \frac{rep_{fac}^{t}}{rep_{fac}^{c}}
\end{multline}
\begin{multline} \label{eq:core}
f_{core} = 0.2 \; cpu_{util} + 0.2 \; mem_{util} + \\
0.3 \; net_{util} + 0.3 \; \frac{rep_{fac}^{t}}{rep_{fac}^{c}}
\end{multline}

IoT-Edge-Processor service is memory intensive while Cassandra service is intensive to all 
resources equally; these sensitivities have been reflected in Eqs~\ref{eq:edge} and~\ref{eq:core}. 
The replication factor of IoT-Edge-Processor service has been set relative to Kafka service and the
replication factor for Cassandra service has been set based on IoT-Edge-Processor service. 
Therefore, as can be inferred, Eqs~\ref{eq:agg}, \ref{eq:edge} and \ref{eq:core} strive to maintain 
the balance between each service and it's lower level service. Note that, replication factor
can be defined differently depending on the managed application logic.

After some time, by adding more and more sensors, the application reaches it's upper
limit capacity; an upper limit capacity for a cloud application may be set for various
reasons~\cite{barna2017delivering}. We let the application to run at full capacity for some time 
and then we configure the client program to remove virtual sensors with the same Poisson 
process to see if Elascale shrinks the whole application accordingly.

Tables~\ref{tab:macro-settings} and~\ref{tab:micro-settings} show the settings for
microservices and macroservices.
\begin{table}[!htb]
 \renewcommand{\arraystretch}{1.3}
  \caption{Macroservices' Settings (VM)}
  \footnotesize
  \centering
   \begin{tabular}{|l||p{2cm}lp{1.3cm}|}
     \toprule
     Layer          &   Service   &   VM (OpenStack)   &  Container    \\
    \midrule\midrule
    Aggregator   &  Kafka   	  &  m1.small  &  Type\_a  \\ 
    Edge-Cloud   &  IoT-Edge-Processor   & m1.medium    & Type\_b \\
    Core-Cloud   &Cassandra & m1.large  &  Type\_c \\
    \bottomrule
  \end{tabular}
  \label{tab:macro-settings}
\end{table}

\begin{table}[!htb]
 \renewcommand{\arraystretch}{1.3}
  \caption{Microservices' Settings (Containers)}
  \footnotesize
  \centering
  \begin{tabular}{|l||rrr|}
   \toprule
    Type           &   Network   &   RAM   &  CPU Quota of VM  \\
    \midrule\midrule
    Type\_a   &  dedicate overlay   &  512\,MB  &  25.0\%  \\ 
    Type\_b   &  dedicate overlay   &  1250\,MB  &  33.0\%  \\
    Type\_c   &  dedicate overlay   &  3\,GB  &  50.0\%  \\
    \bottomrule
  \end{tabular}
  \label{tab:micro-settings}
\end{table}

%-------------------------------------------------------------------------------------
\subsection{Results and Discussion} \label{res-dis}
\begin{figure*}[!t]
\includegraphics[width=\textwidth]{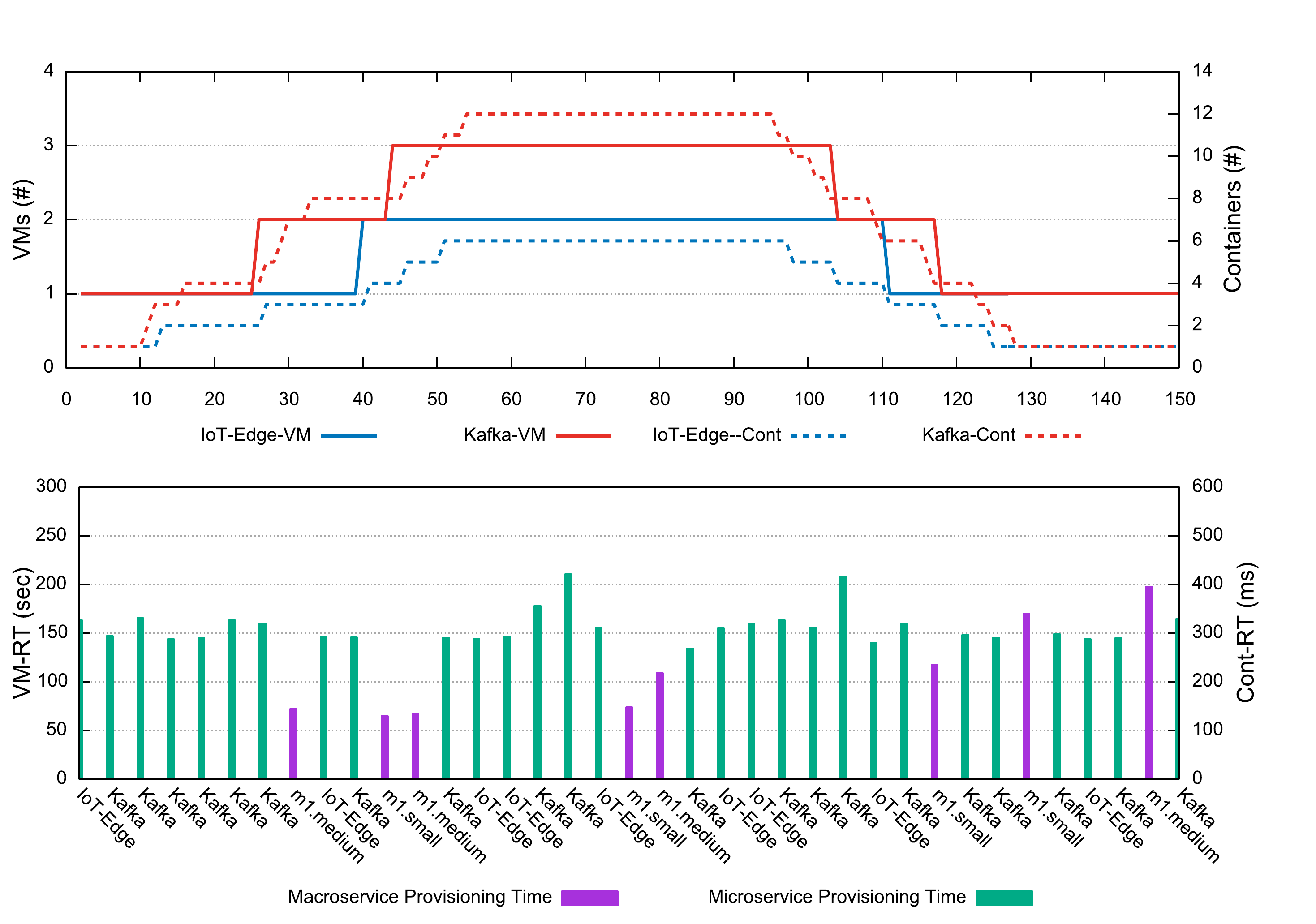}
\caption{Auto-scaling at both micro and macro services (top) and provisioning time for both type of services (bottom). As can be seen in the top plot, the replication factors introduced in reactive formulas 
(i.e., Eqs~\ref{eq:agg}, \ref{eq:edge} and \ref{eq:core}) have contributed to the scaling process. 
As we only turned scalability on for the Edge services, the core services have not been shown in the graph.}
\label{fig:res}
\end{figure*}

We set the upper capacity limit for our application as Table~\ref{tab:limit}. Elascale has
been deployed on the Swarm Manager node using the same command presented in 
Listing~\ref{lst:elascale}. The top panel in Figure~\ref{fig:res} shows the number of containers 
and VMs for the sample IoT application during the experiment. The solid lines represent the number of 
VMs and dashed lines show the number of containers in the application during the experiment 
that took around 150 minutes; we refer to 
each minute of the experiment as an iteration. For the first 10 minutes, application is working with 
initial sensors so no scaling has been initiated. It can be seen that initial 
configuration for the edge is one virtual-container-sensor, one Kafka container, 
and one IoT-Edge-Processor container (see the first 10 minute in Figure~\ref{fig:res}). 
\begin{table}[!htb]
 \renewcommand{\arraystretch}{1.3}
  \caption{Upper Capacity limit for the Application}
  \footnotesize
  \centering
  \begin{tabular}{|l||ccc|}
    \toprule
    Service           &   VM (\#)   &  Container per VM    &  Container (\#)  \\
    \midrule\midrule
    Kafka   &  12   &  4  &  48  \\ 
    IoT-Edge-Processor   &  8   &  3  &  24  \\
    Cassandra   &  1   &  3  &  3  \\
    Visualization   &  1   &  1  &  1  \\
    \bottomrule
  \end{tabular}
  \label{tab:limit}
\end{table}

At iteration 9, we turn the client program on to request new sensors. For any request, the application 
provisions a virtual-container-sensor, that includes 3 virtual probe sensors. As can bee seen in the top plot of 
Figure~\ref{fig:res}, Elascale first scales the Kafka microservice by adding more 
containers as the 70\% threshold has been reached according to Eq~\ref{eq:agg}. Consequently, 
after some time, around iteration 13, Elascale scales the IoT-Edge-Processor service as well. The scaling
out process of microservices is going on until around iteration 25 in which the Elascale, this time, 
scales the macroservice, i.e., adding one VM, for the aggregator service as the existing VM 
is filled with containers. We can see that the macroservices scaling is also happened for the
IoT-Edge-Processor service at iteration 40. This cascading scaling continues until application reaches
its capacity limit around iteration 55. Afterwards, requests for new sensors will be rejected 
by the application. 

\begin{table*}[!t]
 \renewcommand{\arraystretch}{1.3}
  \caption{Active auto-scaling solutions in the open source community. A solution can be used `as-a-service'
  if it can be relatively employed in a simple and straightforward manner. 
  The `Extendability' refers to the amount of 
  required time and effort to extend the solution for other technologies or use case scenarios.}
  \label{tab:autoscale}
  \begin{tabular}{|l||c|c|c|c|c|c|}
    \toprule
    Name & Microservice & Macroservice & Application Agnostic & Cloud Agnostic & as-a-service & Extendability\\
    \midrule\midrule
    Ladder~\cite{ladder} & No & Yes & N/A & Could be & Could be & High \\
    Autoscaler~\cite{autoscaler} & No & Yes & N/A & Yes & No & Medium \\
    Orbiter~\cite{orbiter} & Yes & Yes & Partially & Partially & Could be & High \\
    Cluster-autoscaler~\cite{proportional} & Yes & No & Yes & N/A & No & Low \\
    App-autoscaler~\cite{app-autoscaler} & No & Yes & Yes & Yes & Yes & Medium \\
    Zenscaler~\citep{zenscaler} & Yes & No & Yes & N/A & No & Medium \\
    k8s-kapacitor-autoscale~\cite{k8s} & Yes & No & Yes & N/A & No & Low \\
    \textbf{Elascale} & Yes & Yes & Yes & Yes & Yes & High \\
    \bottomrule
  \end{tabular}
\end{table*}

Around iteration 100, we set the client program to remove the virtual sensors with the same 
process. As can be seen, Elascale shrinks both microservices and macroservices to maintain
optimized resource utilization. Around iteration 130, Elascale scales-in the application 
to the initial state as all the added sensors have been removed by the client 
program. The lower threshold for in-scaling has been set to 40\% for all services. 

As the application owner, we set the ``auto\_scale'' parameter for Cassandra micro
and macro services to ``False'' as based on our experiments, it wouldn't be beneficial 
to scale Cassandra datastore at high load. Adding more nodes or storage capacity to 
Cassandra datastore at high load, will have negative effects on application performance 
for a long time (even hours) due to data replication and synchronization processes in  
background.

The bottom plot in Figure~\ref{fig:res} shows the provisioning time for both macroservices
and microservices. In macroservice level, provisioning means a) creating the VM at backend cloud
b) installation of Docker services, c) joining to the application swarm cluster and c) labeling 
node based on their roles in the application. As can be seen in the bottom plot (i.e., blue bars), 
provisioning macroservices takes 50 to 150 seconds depending on the VM specifications. In terms of
microservice, the provisioning time is in order of milliseconds. Provisioning at microservice
includes, loading the Docker image (images will be available locally after first instantiations)
and configure it to be part of the target service. 

It worth noting that provisioning time is different than ``\emph{contribution time}''. 
We refer to contribution time as the amount of time that is needed for the new resources 
(i.e., VMs or containers) to virtually contribute into the application. 

Contribution time is based on the application logic and the nature of the service that is being
scaled. As a general rule, stateless services (e.g., load balancers) have a contribution time 
close to their provisioning time, while statefull services have a much longer contribution time 
compared to the provisioning time (i.e., distributed datastores). As a result, elasticity 
can be quantified based on provisioning time while scalability is more related to 
contribution time~\cite{khazaei-ficloud-2017}.

%==========================================================
\section{Related Work} \label{rw} 
Autoscaling has received significant attention from academia and industry, especially with the emphasis 
placed on sustainable computing and the emergence of cloud computing and its elastic resources. 
Most approaches to autoscaling are application/platform and infrastructure specific, and rely on 
expert knowledge of the application, or on exhaustive experimentations to derive such knowledge. 
Provided that, reactive and predictive solutions are proposed for autoscaling; a survey of such 
existing solutions can be found in~\cite{lorido2014review}. Such solutions are orthogonal to our 
solution and are not discussed here.
\balance

Leveraging both macroservices and microservices is a promising approach to develop and deploy
cloud applications as each technology brings different features to the table
~\cite{barna2017delivering, khazaei2016efficiency} for software engineers and application providers. 
Both VMs and containers
can be directly controlled and managed by the application in an autonomous manner to maintain 
the SLAs as well as operation costs. However, most of solutions proposed in the literature, are 
either focused on application (i.e., containers) or infrastructure (i.e., VMs). As mentioned above,
most of them are also application or cloud specific. There is no solution out there to be first
application and cloud agnostic and second considers both macro and microservices at the same time.

In open source community there has been some efforts to provide general purpose solutions and 
frameworks to address and provide auto-scalability and monitoring as service though, none of 
which is truly application and cloud agnostic while considering both micro and macro services.
Table~\ref{tab:autoscale} summarized active projects in open source community to date. 
We also pair Elascale in this table to facilitate a straightforward comparison.

In Table~\ref{tab:autoscale} ``Extendability'' refers to the potential of the solution for supporting
other type of micro or macro services; the amount of development effort is the primary criteria in
this quality. As can be seen, Elascale can scale both containers and VMs regardless of the application
or cloud service provider.

%==========================================================
\section{Conclusions} \label{con}
In this paper we presented and evaluated Elascale that is a cloud/application agnostic auto-scaling engine and 
monitoring system. Elascale itself may be deployed as a microservice on the application cluster manager node
and establishes monitoring and auto-scalability automatically. More specifically, it first discovers
microservices and macroservices then incorporates the user inputs regarding final customization through a web
UI and finally monitors the whole application stack to be scaled in or out if deemed necessary.
Also, Elascale provides a monitoring dashboard, in which the application owner can see the live status of
the whole application stack. Elascale has been designed to be highly extendable, incorporating
new scaling algorithms in particular.

As future work, we plan to add Kubernetes support and a generic-sophisticated predictive scaling algorithms 
to Elascale for various types of cloud applications.

\begin{acks}
This research was supported the Natural Sciences and Engineering Council of Canada (NSERC), 
and the Ontario Research Fund for Research Excellence under the Connected Vehicles and 
Smart Transportation (CVST) project.
\end{acks}